# Giant Hub Src and Syk Tyrosine Kinase Thermodynamic Profiles Recapitulate Evolution

J. C. Phillips

Dept. of Physics and Astronomy, Rutgers University, Piscataway, N. J., 08854

## Abstract

Thermodynamic scaling theory, previously applied mainly to small proteins, here analyzes quantitative evolution of the titled functional network giant hub enzymes. The broad domain structure identified homologically is confirmed hydropathically using amino acid sequences only. The most surprising results concern the evolution of the tyrosine kinase globular surface roughness from avian to mammals, which is first order, compared to the evolution within mammals from rodents to humans, which is second order. The mystery of the unique amide terminal region of proto oncogene tyrosine protein kinase is resolved by the discovery there of a septad targeting cluster, which is paralleled by an octad catalytic cluster in tyrosine kinase in humans and a few other species. These results, which go far towards explaining why these proteins are among the largest giant hubs in protein interaction networks, use no adjustable parameters.

Because of the complexity of globular proteins, their evolution is a challenging problem, with only a few quantitative successes so far, notably for small proteins, Hen Egg White (lysozyme $c$, ~ 140 amino acids, aa) [1] and neuroglobin (~150 aa), using thermodynamic scaling theory [2]. Here we discuss much larger nonreceptor (no transmembrane domain) tyrosine kinase Syk (635 aa) and proto-oncogene tyrosine protein kinase Src (539 aa), which are giant hub signaling enzymes that can transfer a phosphate group from ATP to a protein in a cell [3,4]. These proteins are well described on their human Web-based Uniprot homepages. Syk is a giant hub in human protein interaction networks, with 7 molecular functions and 57 biological processes (Uniprot), while a more recent search [5] listed Syk in its Table 2 as the most connected protein localized in both the cytoplasm and cell membrane (CMP) with 105 interactions. The corresponding numbers for Src, a mitochondrion CMP, are 114 interactions on the Uniprot list, and 208 interactions by the keyword methods of [5].   Standard methods based on aa sequence conservation and structural similarities [6,7] enable division of these large protein into three domains and two transition regions, but the changes in their domain structures with evolution are small and not easily quantified.

Progress in analyzing protein evolution now relies on thermodynamic scales, which are of several types.   Here we use both classical and modern hydropathicity scales $\Psi$(aa), which describe the in/out globular folding of protein chains by hydrophobic (in) and hydrophilic (out) aa interactions.   The aa-specific classical scales, especially the standard KD scale based on enthalpic changes of short (< 7aa) synthetic peptides from water to air, are well suited to describing protein functions that are primarily first-order (a few large interactions) [8].   The modern ultraprecise MZ scale, based on critical points and fractals, corresponds to sequence segments of order a membrane thickness (~ 20 aa) or longer [9], and describes protein functions and changes that are primarily second-order (many small interactions) [9,1].   See Fig. 1 for a comparison of the two cales b applied to Syk profiles.

Multiple functions and processes are well suited to analysis by statistical and thermodynamic methods, with emphasis on multiple length scales W.   Here W is the width of a sliding rectangular window, and $\Psi$(aa,W) is a the smoothed hydropathicity of a segment of length W = 2N + 1, which is centered on aa, and has wings of length N aa on each side. To orient ourselves, we show in Fig. 1 the hydroprofiles $\Psi$(aa,21) of human Syk  using both the first-order KD scale, and the second-order MZ scale.   Careful study of these profiles, especially the hydrophobic peaks (spines) [7], and hydrophilic minima which function as elastic hinges [2], shows good correspondence with the Uniprot indicated Euclidean domain and region edges, determined entirely from static structural data.   As expected, the differences between the two profiles are small, and appear to be too small to be quantified by evolution.  Note that most of the differences are concentrated in the central functional domain SH2 and its A and B regional wings.

Because the hydropathic extrema function as spines and hinges, we can quantify their strength by calculating the variances of the profiles:

$$\text{Var}(\Psi(aa,W)) = \Sigma \ ((\Psi(aa,W) - <\Psi(aa,W)>)^2 = \Sigma \ \Psi(aa,W)^2 - n(<\Psi(aa,W)>)^2 \qquad (1)$$

With increasing W, smoothing reduces all variances, but this common factor can be removed by taking variance ratios of species pairs. This procedure was highly effective for the 140aa protein lysozyme $c$ [1], and it might work well for Syk and its functionally active 182 aa domain SH2-2. In fact, it works far better than one might have expected from Fig. 2. The Human/Chicken ratio functions Variance [$\Psi$(aa,W, MZ)] and [$\Psi$(aa,W, KD)] are shown in Fig. 3.

We can interpret the Fig. 3 variance ratios as reflecting the differential human/chick changes in the large (peaks) and small (hills) knobs in the protein globular surface, as measured thermodynamically in terms of first-order or second-order conformational changes. Below comparisons show that here the source of the differences is first-order in Chick, so the broad peak in the Human/Chick ratio corresponds to a smoother Chick globule in the SH2 domain. The satellite variance peaks are displaced from the valley at the center, and show additional smoothing. Their displacement from the center of ~ 20 aa is the typical length for transmembrane protein receptors, with which SYK interacts. The lower/higher W peaks could correspond to shorter/longer correlation lengths at membrane receptor peak/hill configurations.

We were surprised to see the target recognition conserved homology SH2 domain, derived from sequence conservation and Euclidean structural correlations [10], confirmed so simply in the hydropathic variance profile length scale W ratios, but there is another surprise. The Human/Mouse ratio functions Variance [$\Psi$(aa,W, MZ)] and [$\Psi$(aa,W, KD)] shown in Fig. 4, look very similar to those for Human/Chicken in Fig. 3, but the MZ, KD curves have exchanged! To understand this, we need do only one more case, Mouse/Chicken (Fig. 5). This recovers Fig. 3, Hum/Chick, and the common factor is Chicken. This confirms our "guess" above, that Chicken lacks the first-order KD peaks, while later species, Mouse and Human, have the first order peaks and evolve mainly through small second-order hills. Similarly, the difference between Mouse and Human in Fig. 4 is explained by evolution of second-order SYK SH3 hills in Humans.

These Figures show that the Syk profiles of intermediate mammals (like Mouse) exhibit intermediate behavior. Other intermediate mammals, such as Squirrel and Rabbit, exhibit similar (and more nearly similar to human) behavior, which lies outside the scope of this article. Here we have shown that kinase evolution (the subject of 25,000 citations annually) can be quantified thermodynamically. The method uses no adjustable parameters, as the $\Psi$(aa) values for the KD and MZ scales are fixed. The results here are consistent with greater mammal than avian longevity, as multiple second order processes are more reversible and more resistant to mutations than a few first-order ones. One could call this stabilization as the protein self-organized critical point [1] is approached evolutionary telescoping. There may be some similarity to the idea that there are two types of hubs in scale-free protein-protein interaction networks [11,12].

We next look at the Src story. The structure of Src [13,14] (PDB 1FMK) consists of the largely conserved protein-tyrosine domain 267-520, which contains its active and phosphorylation sites between large and small lobes, as in Syk and other kinases, as well as the tandem SH domains SH3 (84-145) and central SH2 (151-248). As in Syk, the central SH2 central domain's function is target recognition. In addition, there is a unique segment (3-83), mentioned in the abstract and text of [13], but not in its Table 1, whose function is mysterious. Evolution of Src from chicken to human is profiled in Fig. 6, using the modern ultraprecise MZ fractal hydropathicity scale [9]. One would expect most of the evolutionary improvements to be concentrated on target recognition (150-250), as in Syk. Instead they are occurring in the mysterious disordered "unique" region 3-83. What is happening here?

The chicken hydroneutral catalytic tetrad 25HHGG28 is turned into a unique 25HGAGGGA31 human hydroneutral superseptad. (The unbiased probability that such a segment, with 7 out of 7 aa being A, G or H, is roughly ~ $10^{-5}$.) The stability of histidine makes it the central and most conserved element of many catalytic triads [15], the most studied examples being Serine-Histidine-Aspartate (chymotrypsin) and Cysteine-Histidine-Aspartate. Catalytic triads form a charge-relay network (central His has pKa = 7), and are excellent examples of convergent evolution [15]. What is most remarkable about the unique Src heptad is that it contains only one catalytic His, while adding hydroneutral Ala and Gly. The three amino acids A,G, and H lie near the centers of approximate classical hydropathicity scales, while they are centered almost exactly by the thermodynamically ultraprecise MZ scale [16].

Replacing the chicken hydroneutral tetrad 25HHGG28 with the 25HGAGGGA31 human hydroneutral superseptad could have two effects. The loss of HH from chicken localizes catalytic effects more in the major protein matrix 267-520, which should improve catalytic specificity [17]. Moreover, the chicken N terminal 3-83 segment is strongly hydrophilic, and it becomes much more stable relative to the C terminal with the human hydroneutral superseptad. Model molecular simulations with self-assembled anisotropic nanoparticles, which have more than two distinct surfaces, each with different properties, suggest that matched ends would be more active [18]. Alternatively one could use the principle of level sets, discussed earlier, to reach the same conclusion through dynamic synchronization [19].

There is some similarity in the evolution of the variance roughness of Src to those of Syk. Src figures similar to Figs. 3-5 are not shown, because their interpretation is less clear. Statistical methods are less accurate when most of the differences are clustered in a small region, such as the human superseptad of Src. Having identified an SH2 septad in Src, we now ask for something similar in Syk SH3, which indeed has an 56AHGRKAHH63 octad! (The unbiased probability that such a segment, with 6 out of 8 aa being A, G or H, is roughly $\sim 10^{-3}$.) However, the presence of three Histidines in this SH3 segment suggests that it supplements allometrically the catalytic activity of the SH2 segment.

The Syk catalytic octad has evolved from hexad GRKVHH in chicken, which hexad has little in common with normal catalytic triads. Thus both Syk and Src have substantially evolved in either respectively their catalytic function or their targeting function. The Syk and Src SH3 clusters are also absent most other mammals, and appear to be rare (the Syk cluster is also present in cats, dogs, horses, baboons and orangutans, but not chimpanzees). It is easy to miss this functional evolution using multiple sequence alignments and structural homologies alone.

Because of the striking appearance of the human Syk catalytic octad in cats, dogs and horses, we can use the giant hub to examine the evolutionary genetic effects of the domestication of these species. This question has been much debated with respect to cat and dog intelligence [20], but here we are asking a different question, the physiological advantages of having a more advanced protein network. We anticipate that dogs are more advanced, because they have been associated with humans for much longer times, but what are the differences between the three species and human Syk molecular structures?

This question is most easily answered with length-dependent variance ratios similar to those shown in Fig. 3 for chick/human. The results for (dog,cat, horse)/human are shown in Fig. 7. As expected, the dog ratio is closest to human, with cat close by. However, the differences are not uniform. Evolution initially matches the central region, W ~ 50, the half-width of SH2 and SH3, so here dog and cat variance ratios are nearly equal. Both the low and high ends are refined later, and there dog is much closer to human. While these results are consistent with the importance of tandem SH domains [14], they are much more detailed. According to Fig. 7, the differences between cat, dog and human are comparatively large at W = 15, but profiles with W =15 similar to Fig. 2 mostly show only very small and widespread differences between cat and dog. (This is what one would expect as the protein approaches evolution-induced criticality.) The largest difference is the deletion of 317AlaVal318 from cat to dog in the central hydrophilic hinge (see Fig. 2), which makes this hinge more flexible in dog (Fig. 7).

Bioinformatically based thermodynamic scaling is often ultraprecise [16]. If a simple, cost-effective cancer biomarker for early cancer detection (ECD) could be developed, it would enable many cancers to be treated effectively by a combination of known and also cost-effective methods; DeVita has called such a biomarker the Holy Grail of Cancer [21]. At present the leading candidate for such a marker appears to be the 15-mer epitopes selected from 400 aa p53, combined with mucin epitopes in the only large-scale (50,000 patients, hence proven cost-effective) clinical studies reported so far [22,23]. The discovered p53 epitopes are more sensitive for ECD than p53 itself, which is more sensitive than any other protein. They also easily distinguish between diagnosable tumors and undiagnosable early cancer onsets. Because it is precise, thermodynamic scaling has confirmed and explained these biomedical results at the fundamental molecular level [24,25].

Thermodynamic scaling is a powerful alternative to molecular methods that use synthetic lattice models as benchmarks [26]. The normal rule (like prefers like), which describes the development of a strong catalytic booster in Syk SH3, has exceptions. Here we found a quite unexpected superhydroneutral booster in the giant hub Src SH3, which has evolved from a catalytic supertetrad interacting with the protein matrix catalytic sites in chickens, to a rare hydroneutral superheptad in humans that enhances target acquisition. There appear to be correlations here, between the two largest kinase interaction networks, and this exceptional hub evolution. The most surprising result, which contradicts all general evolutionary theories, concerns the appearance of the human Syk catalytic hub in some (but not all) other primates, and in cats, dogs and horses (but not bovine or pig) as well. It appears that SH3 can have either the Syk catalytic booster, or the Src target center, or nothing. Given sufficient environmental stability, such as cats, dogs and horses have found with humans, it is possible for domesticated non-primates to develop the same Syk catalytic booster as humans in a few tens of thousands of years, surpassing chimpanzees (few fossil records, so only estimated at 1-2 million years old [27]).

The calculations described in this paper used only an EXCEL Macro built by Niels Voorhoeve and refined by Douglass C. Allan. Our basic assumption that proteins are nearly perfect and are near a self-organized critical point has been proved in a recent study of 7678 NMR data of Human proteins with 10 different structures [28]. The conclusion is that all studied proteins were thermodynamically critical, in the sense that the effective correlation length spanned the entire chain. The present results for grouping human, baboon and orangutans together, and separate from chimpanzee, differ from the results of a global phylogeny which groups humans and chimpanzees together, close to orangutans, and furthest separate from baboon [27,29].

# References

1.  Phillips JC Fractals and Self-Organized Criticality in Proteins.  Phys. A **415**, 440-448 (2014).

2.  Sachdeva V, Phillips JC Oxygen Channels and Fractal Wave-Particle Duality in the Evolution of Myoglobin and Neuroglobin.

3.  Mocsai A, Ruland J, Tybulewicz VL.  The SYK tyrosine kinase: a crucial player in diverse biological functions. Nat. Rev. Immun. **10**, 387-402 (2010).

4.  Hartwell LH,  Kastan MB Cell-cycle control and cancer.  Science **266**, 1821-1828 (1994).

5.  Ota M,  Gonja H, Koike R, et al.  Multiple-Localization and Hub Proteins. PLOS ONE **11**, e0156455 (2016).

6.  Bhattacharyya RP, Remenyi A, Yeh B, et al.  Domains, motifs, and scaffolds: The role of modular interactions in the evolution and wiring of cell signaling circuits. Ann. Rev. Biochem. **75**, 655-680 (2006).

7.  Kornev, Alexandr P.; Taylor, Susan S.  Dynamics-Driven Allostery in Protein Kinases. Trends Biochem. Sci. **40**, 628-647 (2015).

8.  Kyte J, Doolittle RF A simple method for displaying the hydropathic character of a protein. J. Mol. Biol. **157,** 105-132 (1982).

9.  Moret MA, Zebende GF Amino acid hydrophobicity and accessible surface area. Phys. Rev. E **75**, 011920 (2007).

10. Kuriyan J, Cowburn D  Modular peptide recognition domains in eukaryotic signaling.  Annu. Rev. Biophys. Biomol. Struct. **26**: 259–88 (1997).

11. Han JDJ, Bertin N, Hao T, et al.  Evidence for dynamically organized modularity in the yeast protein-protein interaction network.  Nature  430, 88-93 (2004).

12. Ollis DL, Cheah E, Cygler DM, et al. (1992) The alpha/beta hydrolase fold. Prot. Eng. **5**, 197-211.

13. Roskoski R Src protein-tyrosine kinase structure and regulation. Biochem. Biophys. Res. Comm. 324, 1155-1164 (2004).

14. Xu WQ, Doshi A, Lei M, et al. Crystal structures of c-Src reveal features of its autoinhibitory mechanism. Mol. Cell 3,629-638 (1999).

15. Buller AR, Townsend CA (2013) Intrinsic evolutionary constraints on protease structure, enzyme acylation, and the identity of the catalytic triad. Proc. Nat. Acad. Sci. (USA).

16. Phillips JC Hidden thermodynamic information in protein amino acid mutation tables. arXiv 1601.03037

17. Songyang Z, Carraway KL, Eck MJ, et al., Catalytic specificity of protein-tyrosine kinases is critical for selective signalling. *Nature*. **373** (6514): 536–9 (1995).

18. Kobayashi Y, Arai N Self-assembly of Janus nanoparticles with a hydrophobic hemisphere in nanotubes. Soft Matter 12, 378-385 (2016).

19. Phillips JC Fractals and Self-Organized Criticality in Anti-Inflammatory Drugs. Physica A **415**, 538-543 (2014).

20. Roth G, Dicke U Evolution of the brain and intelligence. Trends Cog. Sci. **9**, 250-257 (2005)

21. DeVita VT, DeVita-Raeburn E, The Death of Cancer. Farrar, Strauss and Giroux, New York, p.287 (2015).

22. Pedersen JW, Gentry-Maharaj A, Fourkala E-O, et al. Early detection of cancer in the general population: a blinded case-control study of p53 autoantibodies in colorectal cancer. Brit. J. Cancer **108**, 107-114 (2013).

23. Pedersen JW, Gentry-Maharaj A, Nostdal, A, et al. Cancer-associated autoantibodies to MUC1 and MUC4 - A blinded case-control study of colorectal cancer in UK collaborative trial of ovarian cancer screening. Int. J. Cancer **134**, 2180-2188 (2014).

24. Phillips JC Autoantibody recognition mechanisms of p53 epitopes. Physica A **451**, 162-170 (2016).

25. Phillips JC Autoantibody recognition mechanisms of MUC1. arXiv 1606.07024 (2016).

26. Prado-Martinez J, Sudmant PH, Kidd JM, et al. Great ape genetic diversity and population history. Nature **499**, 471-475 (2013).

27. Peleg O, Choi JM, Shakhnovich EI Evolution of Specificity in Protein-Protein Interactions. Biophys. J. **107**, 1686-1696 (2014).

28. Tang Q-Y, Zhang YY, Wang J, et al. (2016) Critical fluctuations in proteins native states. arXiv 1601.03420.

29. Perelman P, Johnson WE, Roos C, et al. A Molecular Phylogeny of Living Primates. PLOS Gen. **7**, e1001342 (2011).

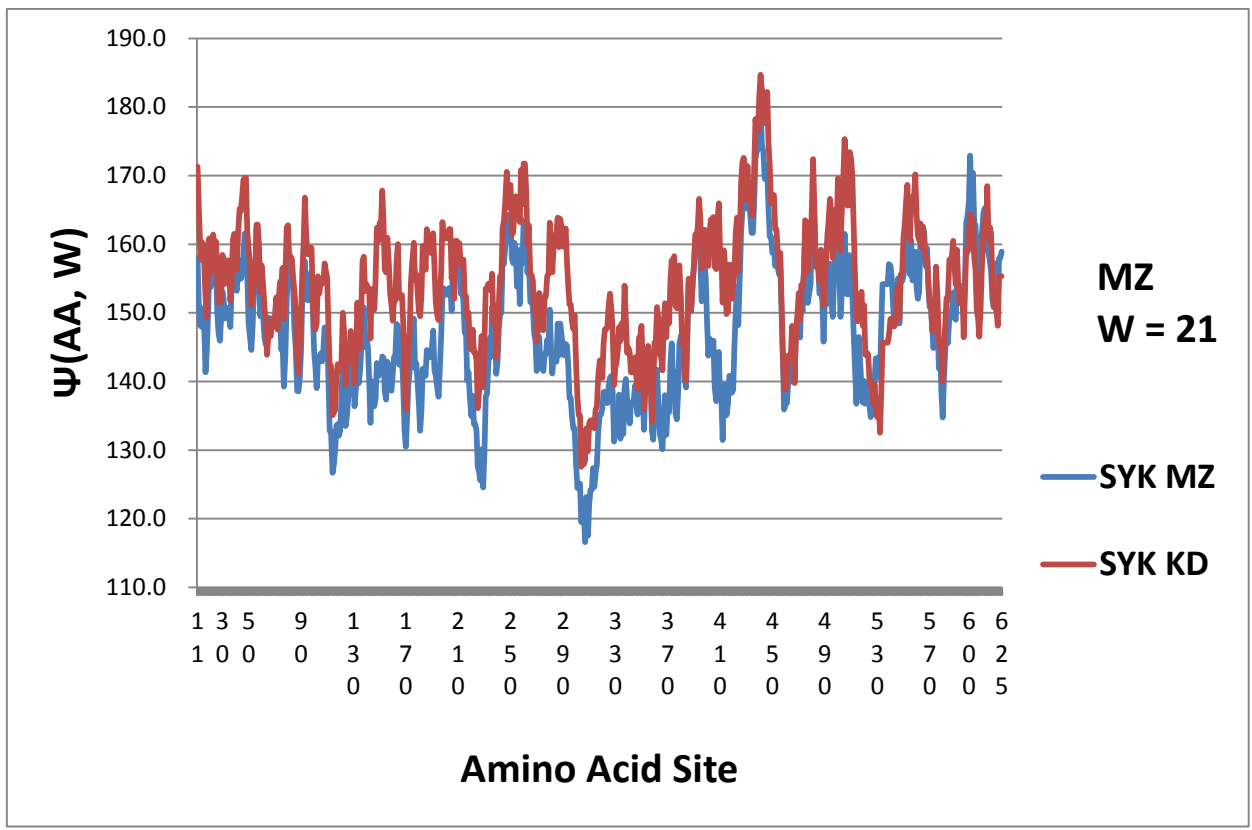

Fig. 1. Hydropathic profiles Ψ(aa,W) for tyrosine kinase Syk using the MZ and KD scales. The value W = 21 corresponds to a membrane length, which is appropriate for signaling globular proteins which interact with transmembrane proteins. The catalytic octad cluster 56-63 is located at the center of SH3 (see annotqations in Fig .2). Note the deep hydrophilic hinge near 300, which reduces coupling between the protein kinase and the tandem SH domains. This coupling is called "clamping" in Fig. 3 of [13]. The hinge is deeper with the MZ scale, which is usually more accurate than the KD scale.

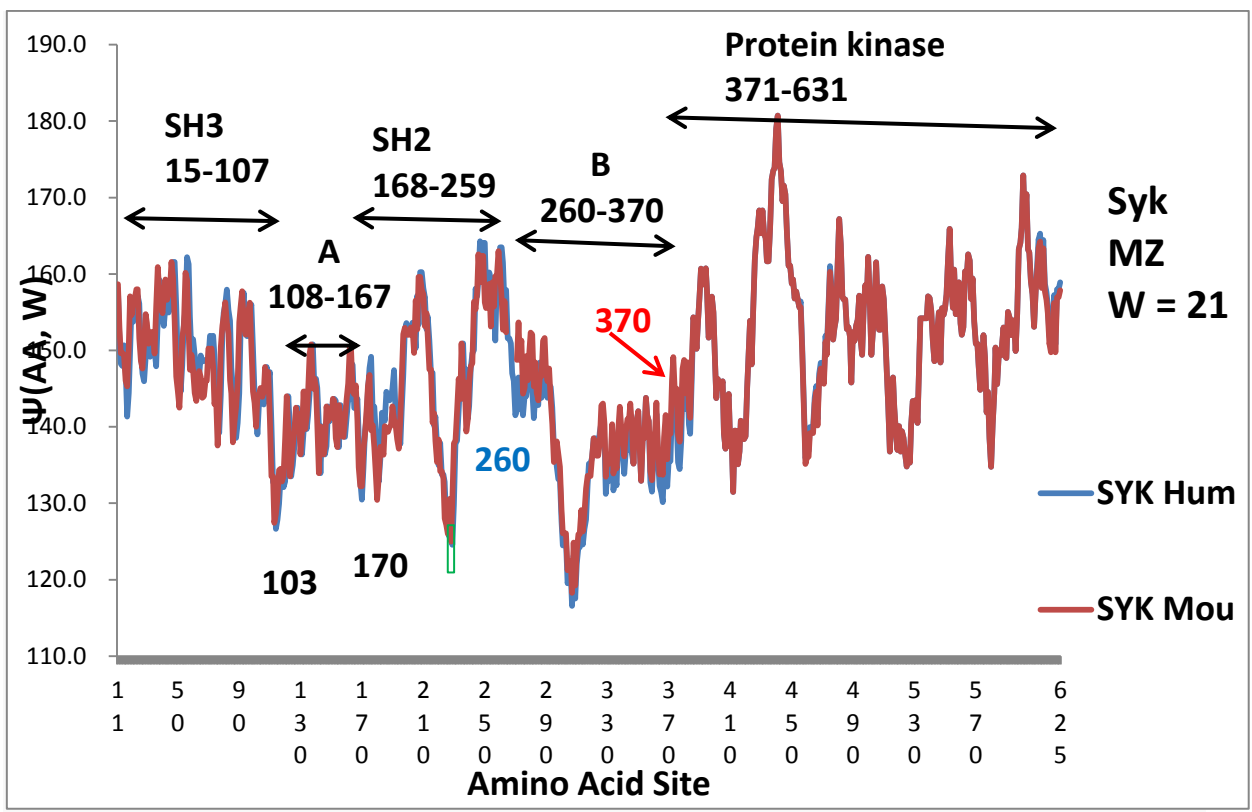

Fig. 2. Hydropathic profiles $\Psi(aa,W)$ for tyrosine kinase Syk for human and mouse using the MZ and KD scales. The annotations are from Uniprot P43405. The value W = 21 corresponds to a membrane length, which is appropriate for signaling globular proteins which interact with transmembrane proteins. The catalytic octad cluster 56-63 is located at the center of SH3. Although the differences are small, their evolutionary significance is clear.

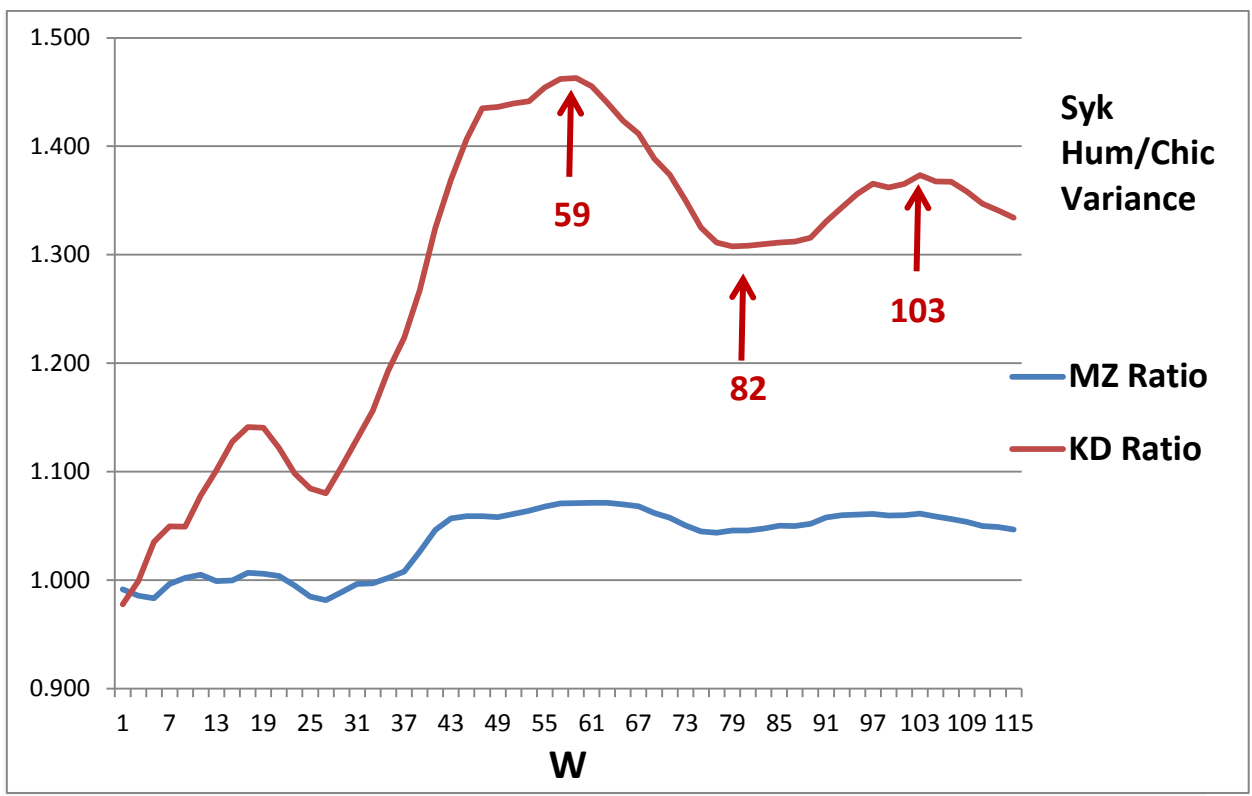

Fig. 3. Hum/Chick globular roughness (variance) ratios for the first-order KD scale and the second-order MZ scale are dominated by a two-peak structure centered on W ~ 82 = width of SH2 domain 259 - 168 (Fig. 1). The two peaks are separated from the center by ~ 20 aa (membrane thickness). That the differences are larger for the KD scale suggests a first-order change in function between chicken and human.

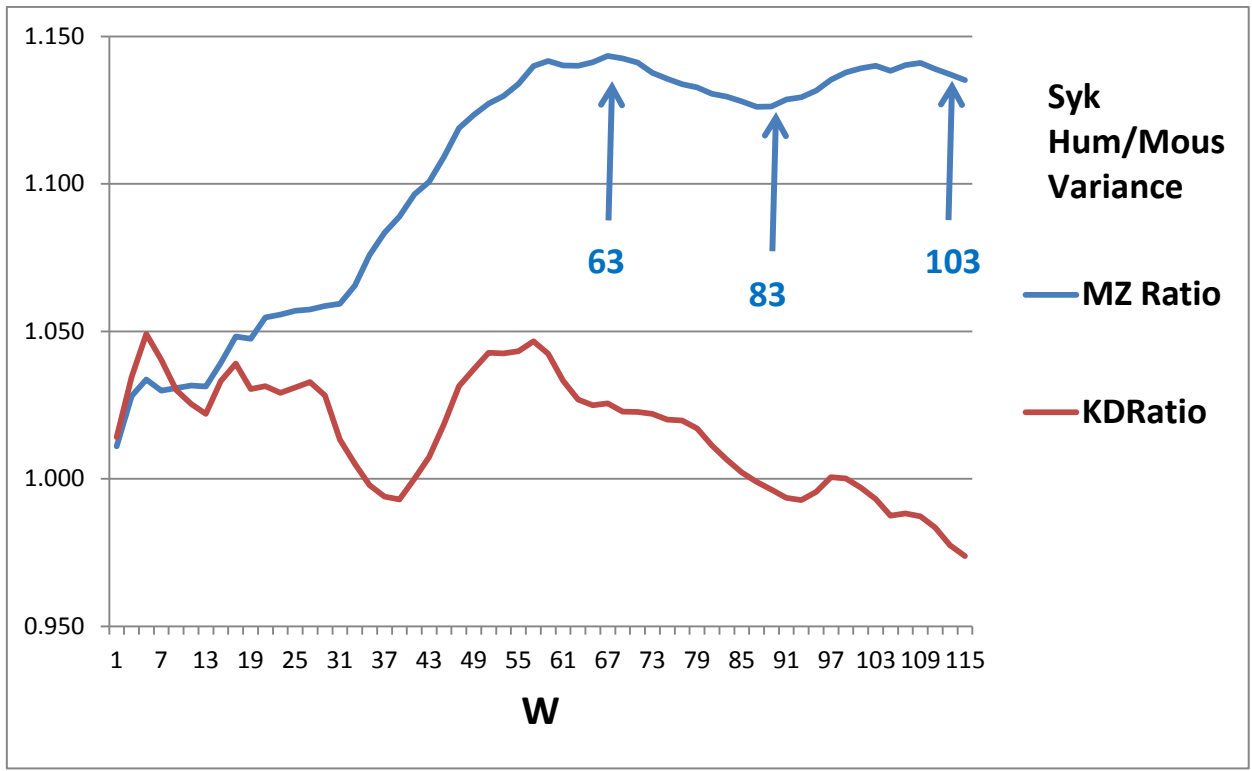

Fig. 4. The Hum/Mouse variance ratio pattern is similar to the Hum/Chick pattern in Fig. 2, except that the curves between the MZ and KD scales have exchanged. In this case, the small differences are better described by the MZ fractal scale.

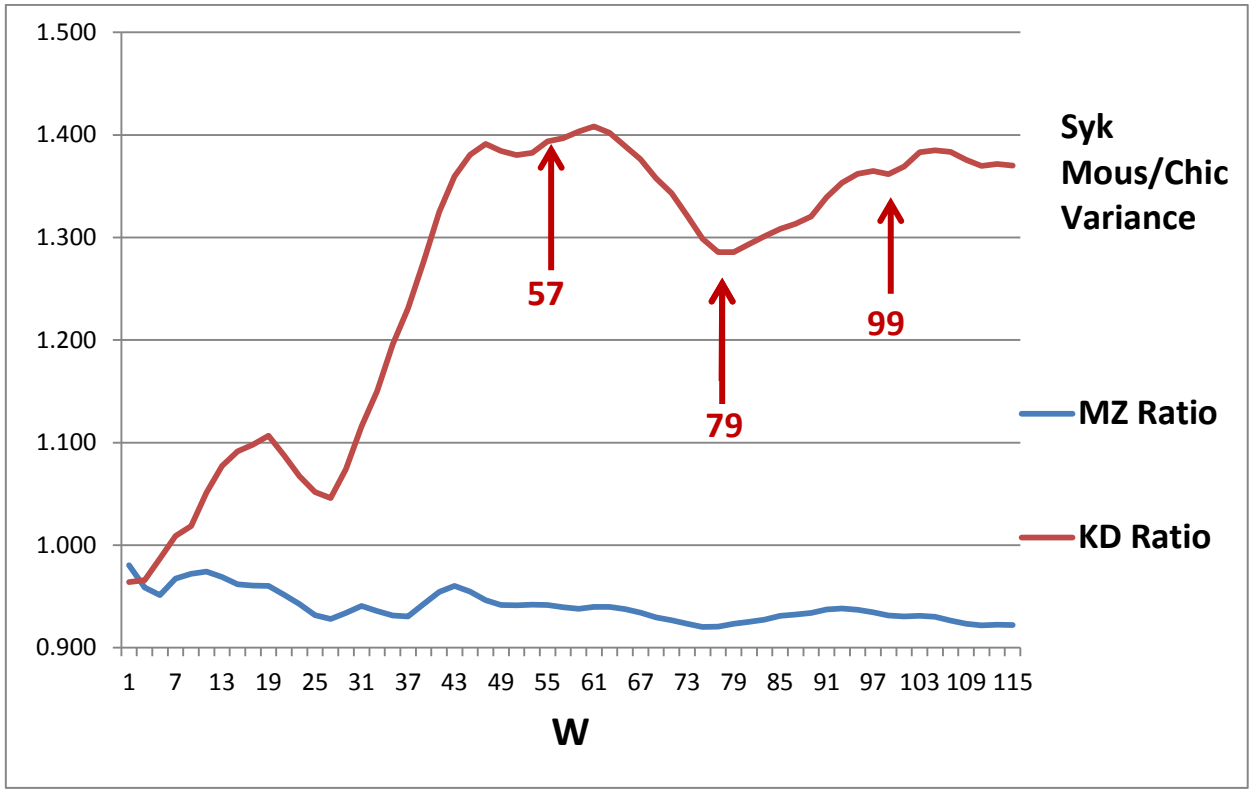

Fig. 5.  The Mouse/Chicken case resembles Fig. 2, Human /Chicken.

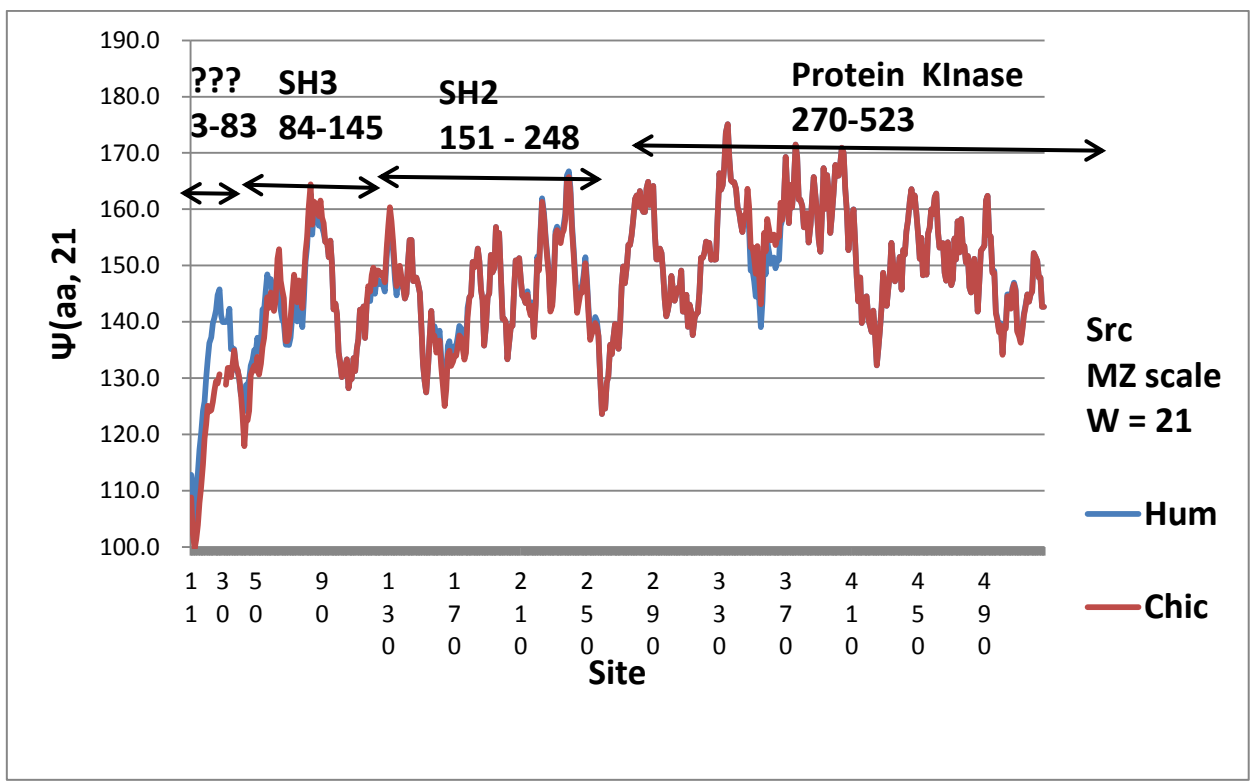

Fig. 6. Hydropathic profiles Ψ(aa,W) for tyrosine kinase Src using the MZ and KD scales. The value W = 21 corresponds to a membrane length, which is appropriate for signaling globular proteins which interact with transmembrane proteins. Note that almost all the evolutionary changes are centered near 30 in the mysterious region near the N terminal. They are associated with the hydroneutral human septad 25-31 discussed in the text.

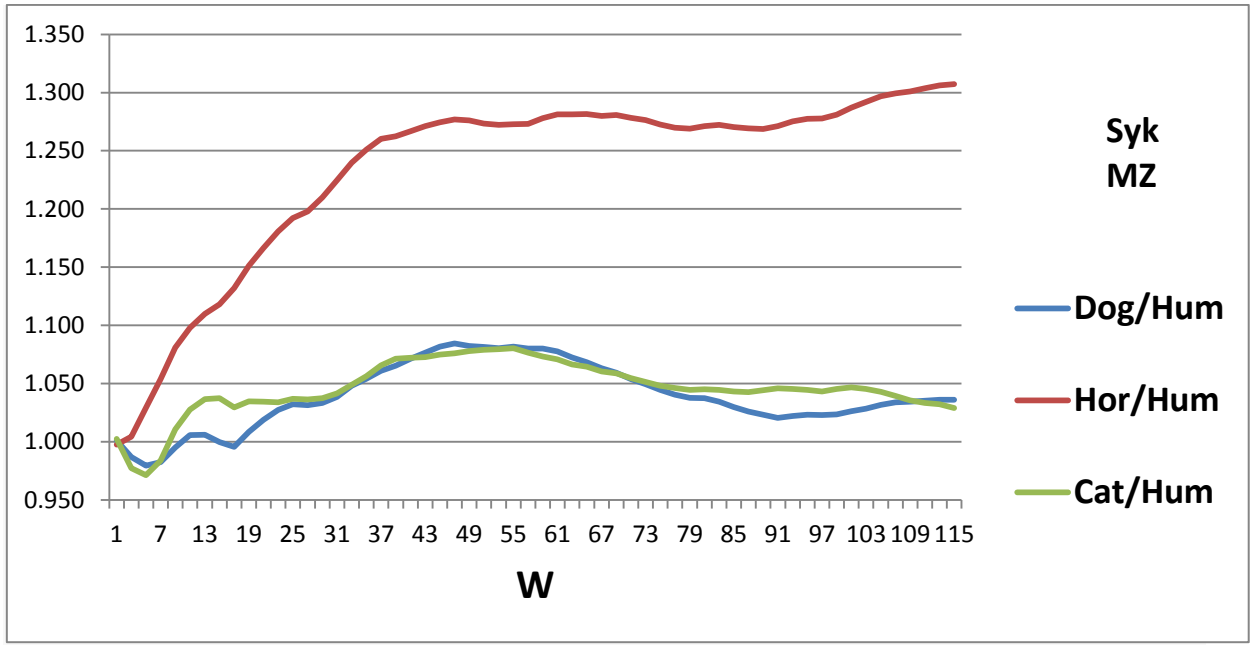

Fig. 7. Comparison of Syk variance ratios for the three domesticated species. Note that dog closely matches human up to W ~ 17. Horse is quite different, presumably because it is less domesticated.

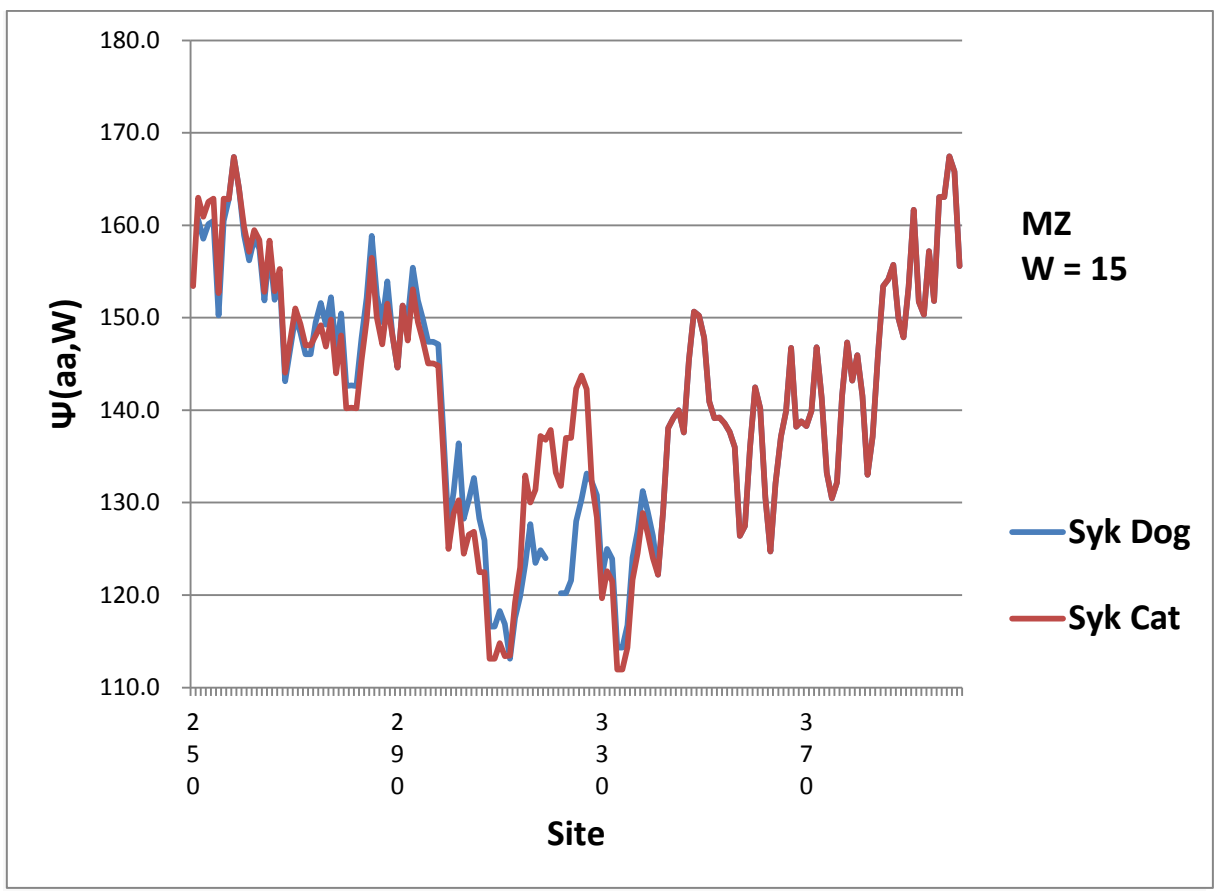

Fig. 8. Details of the central "clamping" hinge [13] in Dog and Cat. The Dog hinge is less hydrophobic near 320, and more flexible. The width of this region is about 15 aa, which is why W = 15 shows a large roughness difference (Fig. 6) between Dog and Cat.